\documentclass[aps,pra,twocolumn,showpacs]{revtex4-1}
\usepackage{braket}
\usepackage{epsfig}
\usepackage{xcolor}
\usepackage{graphicx}
\usepackage{amsmath}
\usepackage{comment}
\usepackage{bm}

\allowdisplaybreaks

\begin{document}

\DeclareGraphicsExtensions{.eps,.EPS,.pdf,.png}

\title{Probing Coherences and Itinerant Magnetism in a Dipolar Lattice Gas}

\author{Thomas Laupr\^etre$^{1,2}$, Ana Maria Rey $^{3,4}$, Laurent Vernac $^{1,2}$, Bruno Laburthe-Tolra $^{2,1}$}

\affiliation{$^{1}$\,Universit\'e Paris 13, Laboratoire de Physique des Lasers, F-93430, Villetaneuse, France\\
$^{2}$\,CNRS, UMR 7538, LPL, F-93430, Villetaneuse, France\\
$^{3}$\,JILA, NIST and Department of Physics, University of Colorado, Boulder, USA\\
$^{4}$\,Center for Theory of Quantum Matter, University of Colorado, Boulder, CO 80309, USA\\
}

\begin{abstract}

We report on the study of itinerant magnetism of lattice-trapped magnetic atoms, driven by magnetic dipole-dipole interactions, in the low-entropy and close-to-unit filling regime. We have used advanced dynamical decoupling techniques to efficiently suppress the sensitivity to magnetic field fluctuations.
 We have thus measured the spin coherence of an itinerant spin 3 Bose dipolar gas throughout a quantum phase transition from a superfluid phase to a Mott insulating phase. In the superfluid phase, a metastable ferromagnetic behavior is observed below a dynamical instability which occurs at lattice depths below the phase transition. In the insulating phase, the thermalization towards a paramagnetic state is driven by an interplay between intersite and superexchange interactions.

\end{abstract}
\date{\today}
\maketitle

{\it Introduction}-
Itinerant magnetism, which relies on the interplay between magnetism and transport in interacting many-body systems, has been the focus of intense investigations tied to superconductivity \cite{Scalapino1995}, quantum magnetism, or quantum transport\cite{Santiago2017, Moriya1984}, in both strongly correlated electrons and atoms. Spinor condensates offer a unique opportunity to revisit this interplay, as ferromagnetism is favored by Bose enhancement \cite{Pasquiou2012}, while Bose-Einstein condensation also strongly favors superfluidity.
However, the connection between superfluidity and ferromagnetism is not trivial \cite{StamperKurn2013} and can be weakened because spin-dependent contact interactions can also favor non-ferromagnetism \cite{Kawaguchi2012,Imambekov2003,Mukerjee2006}.

Furthermore,  long-range interactions between magnetic atoms \cite{Chomaz2023}, heteronuclear molecules \cite{Carroll2024oog}, or Rydberg atoms \cite{Browaeys2020mbp}, can lead to  quantum frustration \cite{Gorshkov2010tos,Yao2018}, or  to complex pattern formations \cite{Defenu2023,Chomaz2023, Su2023} making the expected behaviors even richer. Magnetic atoms are ideal for studying emerging quantum behaviors since,  in contrast to others,  they can be prepared at high filling fractions in optical lattices. Their intrinsically weaker interaction strength, compared to molecules or Rydberg atoms, can be compensated by the use of short-period lattices \cite{dePaz2013, Lepoutre2019, Patscheider2020, Su2023,Douglas2024}.

\begin{figure*}[tb]
\centering
\includegraphics[width= 6.5 in]{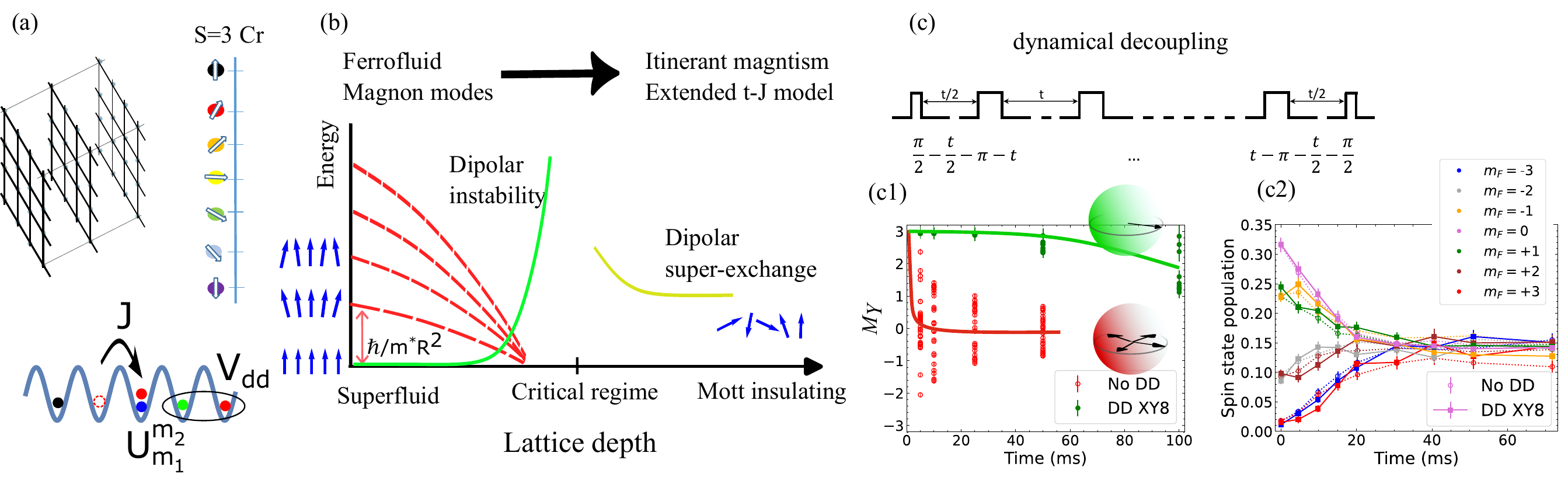}
\caption{\setlength{\baselineskip}{6pt} {\protect\scriptsize (a) We study large spin $S=3$ Cr atoms; with seven  internal sublevels, loaded in a 3D anistropic optical lattice. Our experiment is sensitive to particle motion (through tunneling matrix element $J$), intersite dipolar interactions (DDIs) $V_{dd}$, and on-site interactions between the various spin components $U_{m_1}^{m_2}$. (b) By changing the lattice depth we  vary the ratio between tunneling and interactions energy and  study   out of equilibrium dynamics through out the superfluid to Mott transition. Our experiment reveals a dynamical instability in the SF regime, due to the dipolar coupling between magnons, and a new effect associated with DDIs in the strongly correlated itinerant regime. (c) This experiment is made possible by a dynamic decoupling technique (DD), where $\pi$ pulses are repeatedly applied to the sample.  (c1) DD considerably increases the coherence  lifetime (green data points) compared to the situation without DD (red circles). Solid lines guide the eyes. Spheres depict the generalized Bloch sphere and arrows illustrate that the spin direction becomes random  without DD. (c2) DD does not significantly affect spin population  dynamics in large depth optical lattices.}}
\label{fig1}
\end{figure*}

Here, we  measure the dynamics of transverse magnetization of bosonic chromium atoms in a 3D optical lattice as the lattice depth is varied across the superfluid (SF) to the Mott Insulator (MI) quantum phase transition. We find that deep in the SF regime, an initially imparted ferromagnetic homogeneous structure survives for a very long time. However, the optical lattice hinders the spin currents that are necessary to maintain ferromagnetism \cite{Lepoutre2018}, and  ferromagnetism is broken due to a dynamical instability for lattice depths as the system approaches the SF-MI transition. We compared our data to a simplified analytical model based on hydrodynamic equations, which explains why for our experimental parameters the metastable ferromagnetic character and the superfluid character are lost at similar lattice depths. We also find that the magnetization decay rate  becomes  remarkably insensitive to lattice depth in the MI regime. However, close to the phase transition we find a non-negligible impact of  on-site DDIs on super-exchange interactions, which break the rotational symmetry  of more standard   superexchange interactions arising  from contact interactions only; for our lattice geometry,  they slightly increase the magnetization decay rate.

{\it Suppressing magnetic field fluctuations and gradients using dynamical decoupling}-
For this study, we had to suppress the extreme sensitivity of our system  to magnetic fields, by using advanced dynamical decoupling techniques \cite{Souza2011,Ezzell2023,Li2023tis,zhou2020}. We thus cancel both the effect of magnetic field gradients which can strongly impact the dynamics in the SF regime, and the effect of magnetic field fluctuations on measurement of spin coherences. This allows us to experimentally study the dynamical reduction of the transverse magnetization as a function of lattice depth throughout a SF to MI transition (which complements the study in \cite{Carroll2024oog} on itinerant fermionic molecules). We thus can focus on the many-body physics that is governed by a generalized extended Bose-Hubbard
model given by:
{\small
\begin{eqnarray}
  \hat{  H} &=& - J \sum_{m,\left<j,j' \right>}  \left( \hat{a}^+_{m,j} \hat  {a}_{m,j'} + h.c. \right) \nonumber \\
   &+& \sum_{j,S_t,m_t} \frac{\hat n_j(\hat n_j-1)}{2}  U^{m_t}_{S_t} \ket{S_t,m_t} \bra{S_t,m_t} \nonumber \\
    &+& \sum_{j,S_t,m_t} \frac{\hat n_j(\hat n_j-1)}{2} U^{m_t,d}_{S_t} \left( \ket{S_t,m_t} \bra{S_t+2,m_t} + h.c. \right) \nonumber \\
    &+& \sum_{i >j} V_{i,j} \left[\hat{S}_i^Z\hat{S}_j^Z - \frac{1}{2} \left( \hat{S}_i^X \hat{S}_j^X + \hat{S}_i^Y \hat{S}_j^Y \right) \right]\label{Eq_1}
\end{eqnarray}} where $J$ is the tunneling matrix element, $\hat {a}_{m,j}$ is the bosonic destruction operator of an  atom in Zeeman state $m$ and site $j$; the sum is taken over nearest-neighbours $\left<j,j' \right>$; the on site-interactions  $U^{m_t}_{S_t}$ involve both contact and DDIs and depend both on the  total  spin  of the molecular potential,  $S_t$, and on the total magnetization of the pair of colliding  particle $m_t$ at site $j$; the  on-site DDIs $U^{m_t,d}_{S_t}$ can also couple two molecular potentials with  total spin angular momentum difference $\Delta S_t =2$; the last term is the inter-site DDI term expressed in terms of the interaction strength  $V_{i,j}=V_{dd} \left( \frac{1-3 \cos ^2 \theta _{ij}}{r_{ij}^3}\right)$, with $V_{dd}= \frac{\mu_0 (g_L \mu_B)^2}{4 \pi}$, $\mu_0$ the magnetic permeability of vacuum, $\mu_B$ the Bohr magneton, $g_L\simeq2$, $r_{i,j}=\left|\vec{r_{ij}}\right|$ with $\vec{r_{ij}}$ the vector linking site $i$ and $j$, and $\theta_{ij}$ the angle between the external magnetic field $\vec{B}$ ($\vec{B}  \parallel Z$) and $\vec{r_{ij}}$ . It is written in terms of  the spin-3 operators $S^{\beta=X,Y,Z}_j=\sum_{m,m'}\hat{a}_{m,j}^\dagger \mathcal{S}^\beta_{m,m'}\hat{a}_{m',j}$, $\mathcal{S}^\beta_{m,m'}$ being the matrix elements of the spin-3 angular momentum operator along the spin projection $\beta$.  Here we vary the lattice depth to cover a broad parameter regime, from $J>U,V$, all the way to  $U,V>J$.

The dynamical decoupling (DD) technique used in this work  is a generalization of the echo technique used in Ref.\cite{Gabardos2020} but adapted to tackle  the  situation when  the magnetic field fluctuates in time. A key requirement of a  DD  sequence  is  to  be able to suppress the sensitivity of atoms to magnetic field fluctuations while preserving the   dipolar interactions. Fortunately, this can be satisfied by the use of   $\pi$ pulses around any direction in Bloch space, since DDIs  (last term in Eq. (\ref {Eq_1})), remain invariant under them $S_1^\epsilon.S_2^\epsilon \rightarrow S_1^\epsilon.S_2^\epsilon$, $\forall \epsilon
 \in (X,Y,Z)$).
 It is also important that the chosen DD technique is relatively insensitive to pulse errors, and that it does not create any mean magnetic field in the rotating basis (see \cite{SuppMat}).

{\it Description of the experiment}-
An ensemble of $N\simeq 10^4$  $^{52}$Cr atoms  is prepared in a Bose-Einstein Condensate (BEC) inside  a crossed dipole trap, with a condensed fraction $>80¨\%$. Atoms are polarized in the minimal Zeeman energy state $m_S=-3$. The $^{52}$Cr BEC  is then adiabatically loaded in an anisotropic 3D optical lattice, engineered with $\lambda_L=532$ nm lasers \cite{Alaoui2024}. The controllable depth of the lattice potential $V(x)=V_0\sin^2(k_L x)$ is given in units of the recoil energy $E_{rec}=\hbar^2 k_L^2/2m$ ($k_L=2\pi/\lambda_L$; $m$ is the mass of a chromium atom).
The spin dynamics is triggered by rotating all spins with the use of a Radio Frequency (RF) $\pi/2$ pulse (X pulse). After this preparation pulse, all spins are oriented  orthogonal to $\vec{B}$. We let the spin system evolve under $\hat{ H}$ for a dark time $t$. Then we apply Stern-Gerlach separation and proceed to collective measurement in the basis set by $\vec{B}$, which yields the seven fractional spin populations $P_{m_S}$ ($\sum_{m_S}P_{m_S}=1)$. We thus can monitor spin dynamics as shown in Panel c2 of Fig~\ref{fig1}. Alternatively, when applying a second $\frac{\pi}{2}$ X pulse at the end of the evolution but {\it before} measurement, the
experiment is a Ramsey interferometer.  In that case the quantity $\sum_{m_S} m_S P_{m_S}$ corresponds to the magnetization in a \textit{rotated basis}, which we call $M_Y$. The phase of the interferogram being highly sensitive to magnetic field fluctuations, we use DD during dynamics, which consists of a repetition of X and Y $\pi$ pulses; the Y pulses have a $\pi/2$ phase difference from the initial X preparation pulse. For perfect DD, $M_Y$ is the transverse (normalized) magnetization, $M_Y=\sum_j \langle \hat{S}^Y_j\rangle /N$, $\hat{S}^Y$ being the collective spin component along Y.

We expect distinct  dynamical interplay between the magnetic field gradients  and interactions depending on the lattice depth. At low $V_0$, we anticipate  no spin dynamics in absence of magnetic field gradients \cite{Lepoutre2018}, whereas for large $V_0$ we expect that spin dynamics is weakly sensitive to them \cite{Lepoutre2019}. These expectations guided us for choosing the DD sequences (see \cite{SuppMat}). We show in Fig~\ref{fig1} the result of the Ramsey interferometer with and without DD using a XY8 DD. Panel (c1), taken in the BEC phase, shows a loss of phase coherence in $\simeq 1$ ms in absence of DD; whereas the XY8 decoupling preserves coherence for many 10's of ms. Then the spin dynamics is frozen (as shown indirectly by the stationary value of $M_Y$) as expected. In contract (Panel c2), the spin dynamics due to intersite DDIs persists in presence of XY8 DD, and is barely distinguishable from that in absence of decoupling pulses.

\begin{figure}
\centering
\includegraphics[width= 3.2 in]{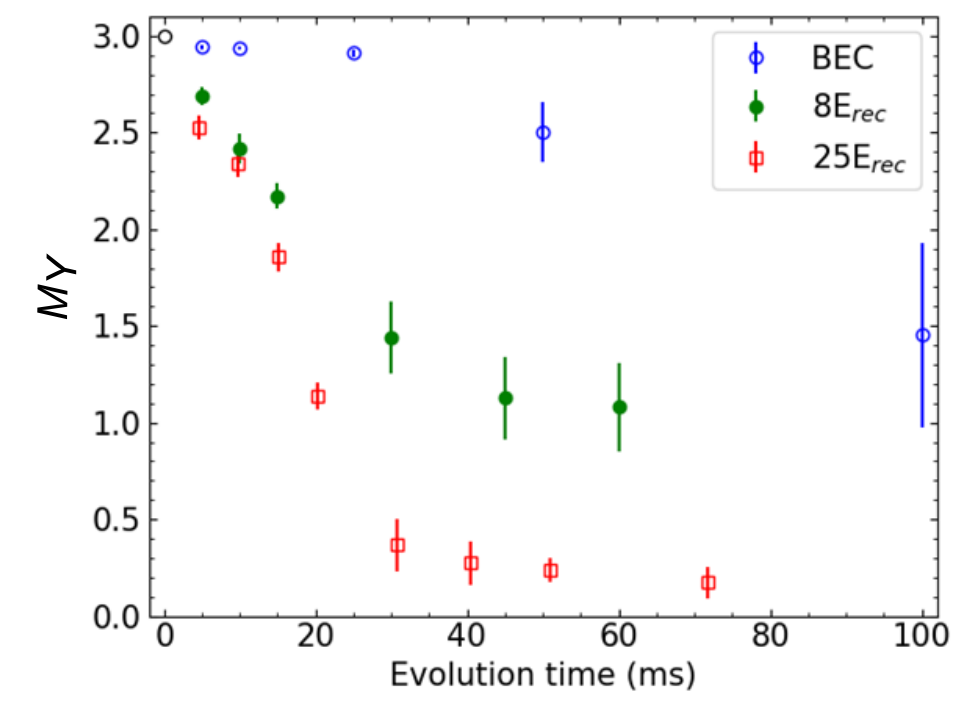}
\caption{\setlength{\baselineskip}{6pt} {\protect\scriptsize Normalized transverse magnetization dynamics for three lattice depths $V_0 = (0, 8, 25) E_R$ respectively (blue) open circles, (green) full circles, and (red) squares. The initial magnetization is 3 for all lattice depths. Error bars account for two standard deviations.} }
\label{fig2}
\end{figure}

{\it Superfluid to Mott transition and ferromagnetism breakdown}-
Thanks to DD, and contrary to previous work \cite{dePaz2016, Fersterer2019}, we can study the dynamics of spin coherence across the SF to MI transition.
Fig.~\ref{fig2} shows the measured  transverse magnetization $M_Y$ as a function of the dark  time for three different values of $V_0$. For the BEC case ($V_0=0$),  the magnetization remains maximal until $\simeq30$ ms, which confirms the metastability of its ferromagnetic character; it ends up decaying at larger times, which we attribute to imperfections of the DD sequence: a mere rotation of the collective spin occurs as more and more RF pulses are applied \cite{SuppMat}. Deep in the MI state ($V_0=25 E_R$), we recover results \cite{SuppMat} of previous study \cite{Gabardos2020}, in which magnetization was deduced from a statistical analysis; here, each experimental point allows for a measurement of the magnetization- which is the direct outcome of the success of the DD approach. Finally, evolution at the SF-MI transition ($V_0\approx 8E_R$ in our setup) provides the first study of coherence in the critical regime.

\begin{figure}
\centering
\includegraphics[width= 3.2 in]{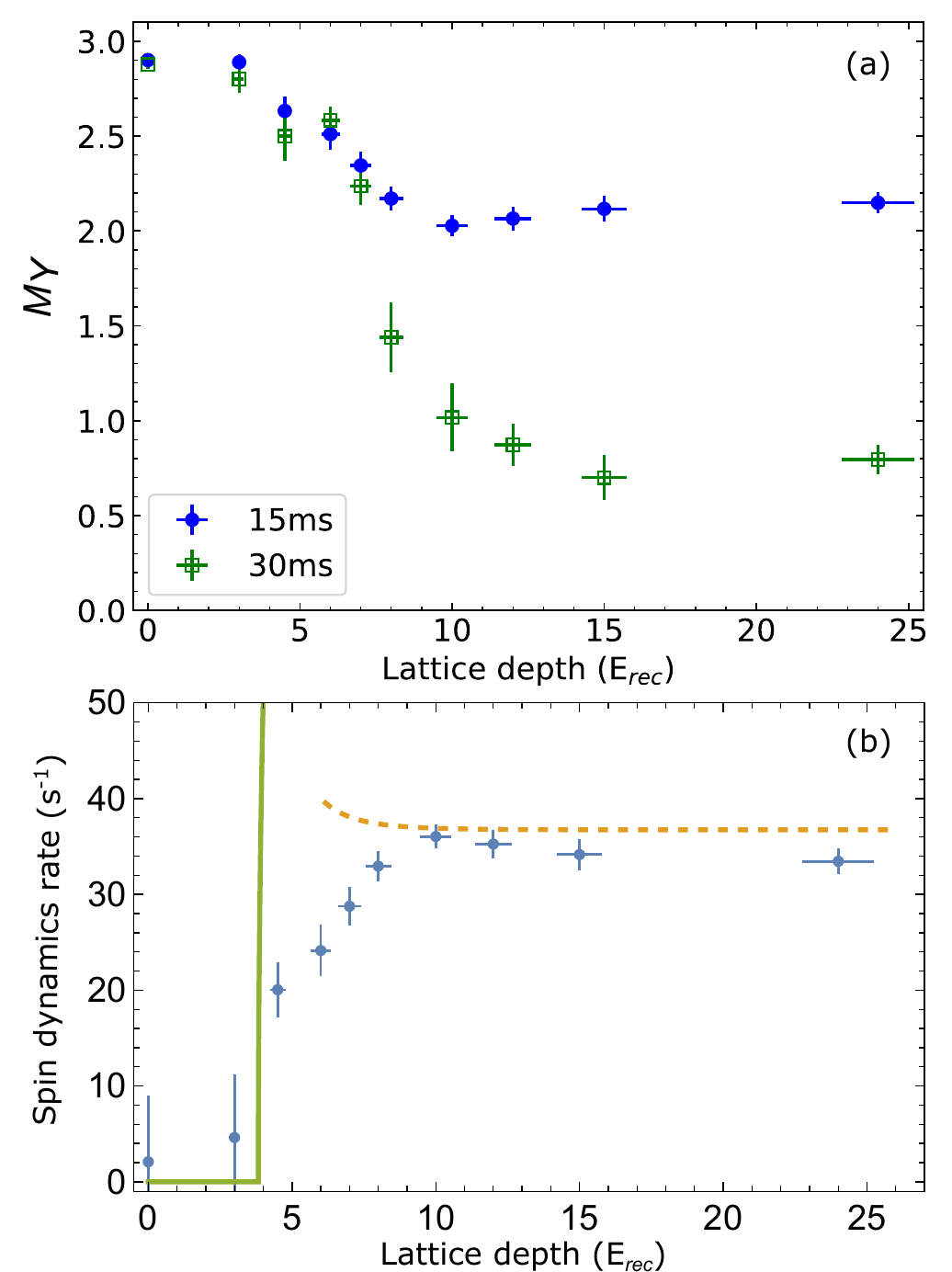}
\caption{\setlength{\baselineskip}{6pt} {\protect\scriptsize  (a) Transverse magnetization as a function of the lattice depth for two dynamical times. (b) Observed and calculated transverse magnetization  decay rate  assuming a quadratic in time decay. The dots show the experimentally extracted rate $\Omega_{exp}$ (see text). The green solid line corresponds to the rate due to the dynamical instability $ \sqrt{2 C_2 \phi_{dd} \omega - 4 \omega^2} $. The yellow dashed line corresponds to the rate deduced by including super-exchange interactions Eq.~(\ref{rateperturbation}).
} }
\label{fig3}
\end{figure}

Fig.~\ref{fig3}a shows $M_Y$ as a function of $V_0$ for two times ($t=15$~ms and $t=30$~ms). The data shows that there is little if any dynamics for $V_0<3 E_R$, but that there is a reduction of magnetization as a function of time for larger lattice depth, still below the SF-MI  transition. Furthermore, spin dynamics is found to be very little sensitive to $V_0$  in the MI regime. Our experimental data thus demonstrate a drastic impact of the transport properties on spin coherences. We quantitatively illustrate this impact on Fig.~\ref{fig3}b where we use data at $t=15$~ms to derive a rate $\Omega_{\rm{exp}}$, assuming a quadratic evolution at short $t$: $M_Y(t)\simeq3 (1- \Omega_{\rm{exp}}^2 t^2)$. Given the extreme complexity to describe spin dynamics in the itinerant regime for our quantum many-body system \cite{Fersterer2019}, we have developed two simplified toy models that capture the main physical mechanisms at play, and allow us  to qualitatively estimate $\Omega_{\rm{exp}}$.

\textit{Superfluid regime}-
 In the BEC case ($V_0=0$) the first $\pi/2$ rotation yields the system in a polarized  state, that is not the lowest energy state since spin-dependent contact interactions favor non-ferromagnetic phases in Cr \cite{Pasquiou2011,Diener2006,Santos2006}. However, in absence of magnetic gradients, ferromagnetism is dynamically preserved, both because a polarized state is an eigenstate of the contact interactions and protected by a \textit{negative} energy gap  \cite{Lepoutre2018}, and because at the mean-field level DDIs introduce no precession when $\left< S_Z \right> = 0$ \cite{Kawaguchi2007csd,Lepoutre2018}. One surprising observation in Fig.~\ref{fig3} is that magnetization is dynamically reduced deep in the SF regime, i.e. significantly below the SF-MI transition.

To explain this surprising result, we performed a stability analysis of the gas, using the hydrodynamic ferrofluid equations that properly account for the gas weak excitations \cite{Lepoutre2018csm}. The validity of this approach stems from the gap protection mentioned above \cite{Lepoutre2018}. In this regime, equations only depend on contact interactions through the system's size, and the local normalized magnetization of the gas $\vec{\mu}$ (with $ |\vec{\mu}|=1$) obeys \cite{Kudo2010heo, Lepoutre2018csm,Farolfi2021qti}:

\begin{equation}
\frac{d \vec{\mu} }{dt}=\vec{\mu} \times \left( \frac{\hbar}{2 m} \left( \vec{a} \vec{\nabla} \right).\vec{\mu} +\frac{\hbar}{2 m} \nabla^2 \vec{\mu}- \frac{g_S \mu_B}{\hbar} \vec{B_{dd}}(\vec{r})
\right)
\label{hydroeq}
\end{equation}
where $\vec{B_{dd}}(\vec{r}_i)=\sum_j V_{i,j} (-\frac{1}{2}\langle \hat {S}^X_j\rangle,-\frac{1}{2}\langle \hat {S}^Y_j\rangle,\langle \hat {S}^Z_j\rangle )$ is the dipolar mean-field, and
$\vec{a}=\vec{\nabla } n / n$ is the relative gradient of the total density $n$. To describe how the optical lattice modifies transport, we replace $m$ in Eq.~(\ref{hydroeq}) by the effective mass $m^*$  which is set by the second derivative of the single-particle dispersion relation. Using a Gaussian ansatz, and in absence of  DDIs, the stationary solutions  are  given by Hermite polynomials $H_n$ that account for a locally polarized gas with spatially varying direction of polarization. The key point is  that DDIs provide a spatially inhomogeneous and non-linear magnetic potential that couples the modes described by $H_n$. Given the (parabolic) saddle-like harmonic nature of the dipolar potential \cite{ODell2004eho}, the originally populated mode with $H_0$ is dynamically coupled to that with $H_2$. For sake of simplicity we use a one-dimensional model, where the dipolar mean-field is set by  $ g_S \mu_B \vec{B}_{dd} (z)= \phi_{dd} (1-z^2/d^2) \left( -\frac{1}{2} \mu_X,-\frac{1}{2} \mu_Y, \mu_Z \right)$.   $\phi_{dd}= \frac{n_0}{3 \hbar}  \mu_0  (g_S \mu_B )^2  S$ \cite{ODell2004eho}, and $n_0$ is the BEC central density. The $(1-z^2/d^2)$ behavior empirically describes the complicated 3D nature of the dipolar mean-field, with $d$ taken as an adjustable parameter.

We look for solutions $\vec{\mu} = \left(f, \sqrt{1-f^2-g^2},g \right) $ with $f= \alpha_0(t) H_0(z) + \alpha_2(t) H_2(z) $ and $g= \beta_0(t) H_0(z) + \beta_2(t) H_2(z) $ with $ \alpha_2(0)= \beta_2(0)=0$, in order to perform a stability analysis. $\beta_0 = \beta_0(0)$ is assumed to be small but not vanishing.  We then find (assuming $(f,g) \ll 1$):

\begin{equation}
    \alpha_2(t)=-\frac{\beta_0 C_1 \phi_{dd} \mathrm{sinh}\left( \sqrt{2 C_2 \phi_{dd} \omega - 4 \omega^2} t \right)}{\sqrt{2 C_2 \phi_{dd} \omega - 4 \omega^2}}
    \label{eqinstab}
\end{equation}
where $\omega = \hbar^2 / m^* R^2$ is the frequency of the first excited mode, $R$ is the cloud's size, and $C_{1,2}$ are defined in \cite{SuppMat}.

For $C_2 \phi_{dd} < 2 \omega$, $\alpha_2(t)$ undergoes small oscillations around the initial value. In contrast when $C_2 \phi_{dd} > 2 \omega$ the mode  grows exponentially with a timescale $\approx (2 C_2 \phi_{dd} \omega - 4 \omega^2)^{-1/2} $, so that any fluctuation in the initial polarization will destabilize the gas, leading to a reduction of the magnetization. The instability criterion $\phi_{dd} > \frac{2}{C_2}\frac{\hbar}{m^* R^2}$ is equivalent to the dynamical instability inferred using the Bogoliubov expansion in \cite{Lepoutre2018csm}.

The main effect of the effective mass is to enable  the criterion $C_2 \phi_{dd} > 2 \omega$ to be satisfied (in practice at $V_0\approx 3 E_R $). Interestingly, the condition $\phi_{dd}^{\xi} \approx \hbar^2/m^* \xi ^2$ defines the distance $\xi$ at which spin domains can be formed in a ferrofluid. Since  in chromium DDIs are not small compared to contact interactions \cite{Chomaz2023}, the ferrofluid character of the gas is totally lost when $\frac{\hbar^2  n^{2/3}}{m^*} \approx \phi_{dd} $ (with $n$ the atomic density), which happens close to the SF-MI transition (where $\frac{\hbar^2  n^{2/3}}{m^*}$ approaches the onsite interaction energy).

We stress that the instability threshold strongly depends on the spatial dependence of the DDIs, characterized by the free parameter $d$. In Fig.~\ref{fig3}b, we have taken $d^2 = 0.55 R^2$, with $R$ the estimated average Thomas Fermi radius of the BEC (to be compared to $0.5R^2$ estimated from \cite{ODell2004eho} in 1D).

\textit{MI  regime}- There are two main effects associated with tunneling in the MI state. The first is the creation of hole-doublon pairs, whose number scales as $J^2/U^2$, where $U$ is the total onsite interaction between two atoms being in a molecular state with a total spin $S_t=6$. Since the Mott transition qualitatively occurs for $U/zJ \approx 5.8$ \cite{Jaksch1998cba}, with $z$ the number of nearest neighbors, spin dynamics is barely modified by the presence of hole-doublon excitations even close to the SF-MI transition (see \cite{SuppMat}).

A second and more substantial effect arises due to on-site DDIs. Because these interactions depend on magnetization and because they couple the different molecular potentials with $\Delta S_t=2$, their impact on super-exchange interactions qualitatively differs from that of contact interactions, which preserve the total spin $S_t$. Using a toy model with two $S=1$ atoms in two sites, we find, up to the second order in time:
\begin{equation}
M_Y=\frac{\langle\hat{S}_Y (t) \rangle}{N} = \frac{\langle\hat{S}_Y(0) \rangle}{N} -\frac{t^2}{2 } \left( \frac{J^4}{8 U_A^2} + \frac{  J^2\bar{V}}{8 U_B} –  \frac{9 V_{\rm{eff}}^2}{8}    \right)
\label{rateperturbation}
\end{equation}
where $U_A$ and $U_B$ depend both on $U_{S_t=0,2}^{m_t}$ (spin-dependent on-site contact interactions, and magnetization-dependent DDIs), and $U_{S_t=0}^{m_t,d}$ which  couples  $S_t=0$ and $S_t=2$; $\bar{V}$ and $V_{\rm eff}$ are defined as the average and quadratic average of DDIs \cite{SuppMat}. In order to provide a qualitative comparison between our experimental data and this toy model, we have evaluated $U_{S_t=0,2}^{m_t},U_{S_t=0}^{m_t,d}$   as a function of $V_0$; we use the $S_t=6$ and $S_t=4$  scattering lengths of Cr atoms  \cite{Pasquiou2010} to evaluate respectively $U_2^{m_t}$ and $U_0^{m_t}$ (or $U_{S_t=0}^{m_t,d}$). We thus find an increase of the rate of spin dynamics due to DDIs by typically 10 percent at the SF-MI transition compared to what is obtained deep in the MI state, which is in qualitative agreement with experimental data.
This reveals an entirely new phenomenon due to on-site DDIs, that can be tuned by the lattice structure (that sets $\bar{V}$) and the shape of the lattice sites (that sets the sign of  $U_B$); spin-dependent contact interactions themselves only modify spin dynamics at the fourth order in time.

\textit{Conclusion}-
In this work we have explored the breakdown of ferromagnetism during the dynamics of a quantum gas due to DDIs.
We find that a metastable ferromagnetic phase becomes unstable when transport is hindered by the use of an optical lattice. Spin coherences associated with ferromagnetism become dynamically unstable in the SF regime when approaching the MI regime, a phenomenon which we expect to also impact other strongly magnetic atoms such as lanthanides \cite{Chomaz2023}. In the MI regime, we find that the dynamical evolution of coherences can be affected by an interplay between intersite and on-site DDIs - which would be even more prominent (and geometry dependent) in the case of heteronuclear molecules and lanthanides. While we could qualitatively interpret our experimental data using relatively simple toy models, our study calls for in depth theoretical investigations of the critical regime. Experimentally, our study was made possible by the use of dynamical decoupling techniques, which provide an extension of the spin coherence time by $\simeq1000$ \cite{SuppMat} and offer a practical alternative to magnetic shielding \cite{Farolfi2019,Farolfi2021}.

\vspace{1cm}

\begin{acknowledgments}
{\it Acknowledgments} We acknowledge financial support from CNRS, Agence Nationale de la Recherche (project DISQuTT - ANR-23-CE47-0016), and QuantERA ERA-NET (MAQS project). AMR is supported by  AFOSR MURI FA9550-21-1-0069, W911NF24-1-0128, and  NSF JILA-PFC PHY-2317149 grants.
\end{acknowledgments}

\newpage

\newpage

{\LARGE Supplemental Material}

\section{I Principle of the experiments}

\begin{figure}
\centering
\includegraphics[width= 3.4 in]{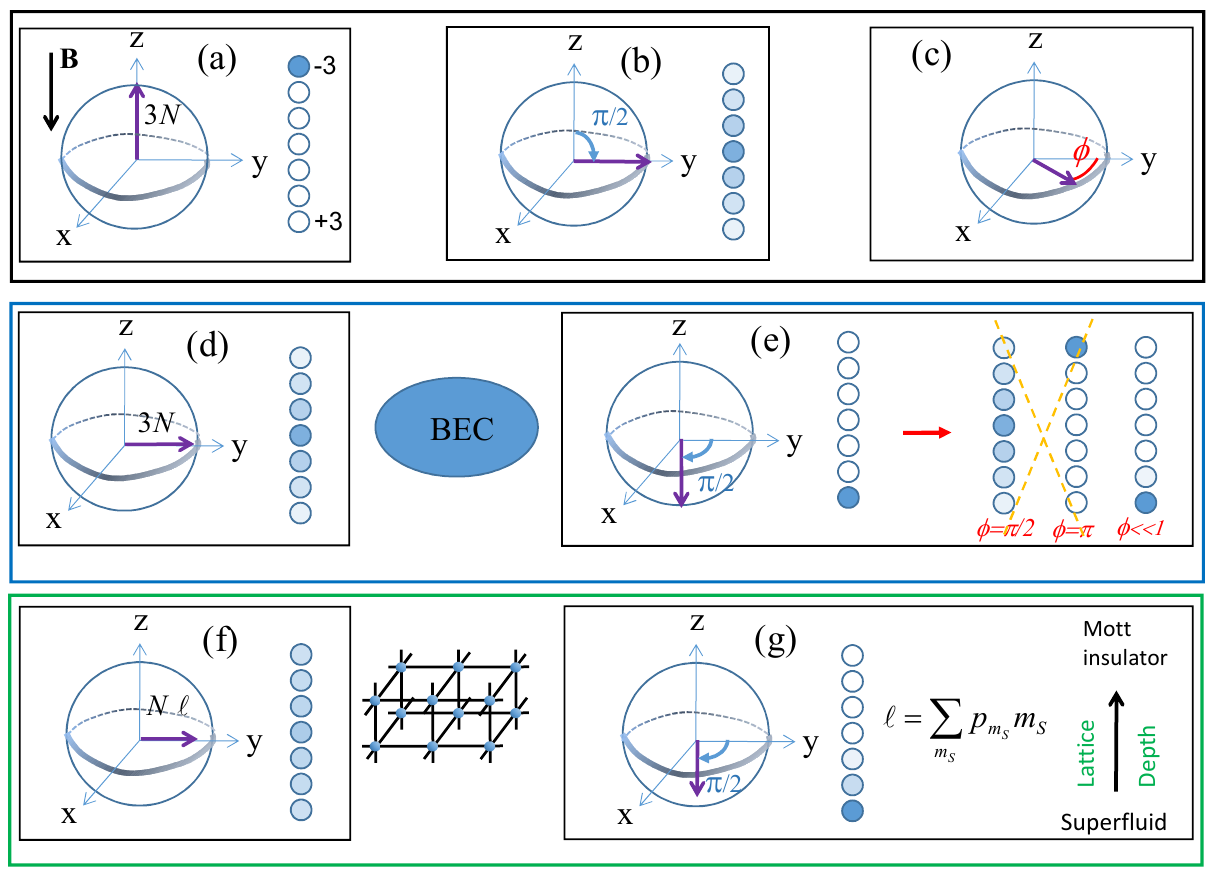}
\caption{\setlength{\baselineskip}{6pt} {\protect\scriptsize Principle of the Experiments. Top: from the ground state (a), spin dynamics is triggered by a $\pi/2$ rotation (b), its tracking is affected by magnetic fluctuations which create desynchronization i.e. uncertainty in the collective spin direction (c). Middle: in a Cr BEC,  Dynamical Decoupling (DD) freezes mean-field spin dynamics (d), which allows to quantify performances of DD in suppressing desynchronization (e). Bottom: deep in the MI phase, DD does not modify quantum spin dynamics (f), and allows to track the the transverse magnetization $M_Y$ through a quantum phase transition (g).}}
\label{figSM1}
\end{figure}

We illustrate in Fig~\ref{figSM1} the principle of our experiments using quantum gases of chromium. Initially, all atoms are in the ground state $m_s=-3$. To initiate dynamics, the spin of the atoms are rotated by a $\pi /2$ RF pulse. During dynamics, the transverse magnetization, initially maximal ($=3N$), can decrease. Its measurement is obtained by performing a second $\pi /2$ pulse at the end of the dynamics, which in practice rotates the measurement basis. To do so, we measure, after this second pulse, fractional spin populations $P_{m_s}$ in the quantization basis set by the external magnetic field ( $\sum_{m_s} P_{m_S}=1)$. This yields $M_Y=\sum_{m_s}m_s P_{m_S}$ which is equal to the (normalized) value of the transverse magnetization ($-3 \leq M_Y \leq +3$).

Due to magnetic field fluctuations, the collective spin of the atoms acquires a random orientation set by the angle $\phi$ in the rotating frame associated with the RF frequency. $\phi$ is the accumulated differential precession angle between the rotating spins and the co-rotating component of the RF field. This technical effect alters the measurement of the magnetization described above.
For example, in absence of spin dynamics, the atomic state remains unchanged, and $M_Y=+3$; after a second $\pi /2$ pulse all atoms should be transferred in the $m_s=+3$ state, but in presence of fluctuations, random values of $\phi$ lead to random values of $M_Y$.

The use of Dynamical Decoupling (DD) technique allows to efficiently suppress the effect of magnetic fluctuations on spin precession. It suppresses as well sensitivity of the systems to magnetic field inhomogeneities (gradients). In the BEC case, and in absence of magnetic gradients, the spin dynamics is expected to remain frozen. Therefore,  the measurement of transverse magnetization with a BEC after a Ramsey sequence can be used to reveal the efficiency of the DD sequence in reducing the effect of magnetic fluctuations: $M_Y=+3$ is then expected for a perfect DD sequence. When the atoms are loaded in a lattice, spin dynamics happens for large enough lattice depths (see main text). Deep in the MI regime, we have checked that DD does not modify spin dynamics as measured in the $Z$ basis (see Fig.\ref{figSM3}); then DD allows to easily track the transverse magnetization dynamics, each experimental realization providing a value for the collective atomic spin.

We have implemented our experimental tests on DD performances either in the SF or deep in the MI regime (see section III below), because for these two extreme cases we knew what to expect ($i.e.$ we knew from past experience and numerical simulations what the spin dynamics should be in absence of magnetic field gradients). In contrast, we did not perform such tests in the critical regime for which we have no benchmark.  Note that the perturbative theoretical model that we have elaborated to assess the robustness of DD to pulse imperfections (see section III below) assumes frozen spins: we do not expect it to be valid in a regime where there is both transport and strong spin-orbit coupling (such as the one provided by magnetic field gradient).
Our goal was indeed to find a regime where DD is insensitive to pulse imperfections; note that it is a separate question than the one on how DD is sensitive to transport.

Finally we indicate characteristics of the quantum gases used in the experiment. We start with a very cold BEC, with a condensate fraction $\simeq80\%$, and a high filling factor when loaded in our lattice potential. In our experimental setup, this leads, for large lattice depth, to a core of doubly occupied sites (doublons), surrounded by a shell of singly-occupied sites (singlons); see \cite{Alaoui2022}.
Starting with typically $N=10000$ atoms, the atomic sample endures losses due to dipolar relaxation \cite{Pasquiou2010}. In the SF regime, these losses are sufficiently slow that the evolution can be considered quasi-static, see \cite{Lepoutre2018}. In the MI regime, losses affect only the doublons, and in $\simeq10$ ms the system is left with only singlons \cite{Alaoui2022}: for $t>10$ ms, the atom number remains stable, $N\simeq 5000$. Therefore, the model that we use for describing the MI regime (see section VI) can safely assume singly occupied sites only, for $t>10$ ms. As in our past work, these singly occupied sites occupy a shell surrounding those sites that have been emptied by dipolar relaxation.

\section{II Experimental parameters for dynamical decoupling}
Our first attempt to reduce magnetic field fluctuations has been to use standard techniques.
Along the $\sim20$ s long total experimental sequence, the beginning of the RF sequence itself is systematically triggered on a slope of the 50 Hz-mains, ensuring repeatability of the associated desynchronization. Besides, a large coil also compensates for the fluctuations detected along the quantization axis and synchronized on the 50 Hz-mains. This contributes to a reduction below 1/10 of these fluctuations and an increase of the total desynchronization time from $\sim1.25$ ms to $\sim2.5$ ms (see Fig \ref{figSM2} (a)). This gain by a factor about 2 was not sufficient to implement our study, which is why we turned towards the use of dynamical decoupling. We give below details on the experimental implementation of our DD sequences. 

The $\frac{\pi}{2}$ RF pulses are $\sim3.6\,\mu$s long, corresponding to exactly 8 oscillations at the Larmor frequency $\omega_\mathrm{Larmor} /2 \pi   \sim 2.22$ MHz in presence of a $\sim0.8$ Gauss field along the quantization direction. This pulse duration is chosen as short as possible to minimize the influence of Larmor frequency fluctuations, while keeping in the linear range of the RF amplifier which is used. The corresponding Rabi frequency is $\Omega_\mathrm{Rabi}/2\pi\sim70$ kHz. The pulse frequency and amplitude are obtained from a Ramsey sequence of a few tens of $\mu$s, short enough that desynchronization is negligible ($\phi \ll 1$). Long-term fluctuations of the Larmor frequency at the $\sim\pm5$ kHz level require to check the pulse frequency regularly during a typical experiment day.

The DD sequences contain a series of $\pi$ pulses regularly positioned in time between the first and second $\pi/2$ pulses. As stated in the main text, these pulses can be either $X$ pulses (which have the same phase as the $\pi/2$ pulses, and correspond to rotations around the X axis), or $Y$ pulses (which have a $\pi/2$ phase shift with respect to the $X$ ones, and correspond to rotations around the Y axis). They all share common values of amplitude and frequency with the  $\pi/2$ pulses, but are twice as long.

The separation $T$ between $\pi$ pulses is 100 $\mu$s, which results from a compromise between minimizing desynchronization in between pulses and reducing effects of pulse distortion over time from a too high repetition rate.

The complete function defining the RF sequence, i.e. the RF amplitude as a function of time, is calculated point by point and reproduced by a commercial Arbitrary Waveform Generator (Keysight 33500B range, AWG) with a typical sample rate $\sim$95 MSa/s, insuring steps small enough to  accurately reproduce the different kind of pulses using embedded interpolation. The signal is then delivered to an excitation coil by a commercial high stability and protected amplifier.

\section{III Comparison between DD sequences, and interpretation}

We have tried various DD sequences in our experiment, see Fig \ref{figSM2} (b). In this section, we first indicate what was their experimental outcome and what is our interpretation. We then propose a model to support our interpretations. We remind the reader that, as explained in  section I, the two figures of merit that we seek for a DD sequence are: (1)- suppression of the effect of magnetic field fluctuations (i.e. desynchronization) on measurement of magnetization, which is characterized in the BEC phase; (2)- no modification of the spin dynamics in the MI regime, at large lattice depth.

\subsection{Experimental protocols and their outcome}

\begin{figure}
\centering
\includegraphics[width= 3.4 in]{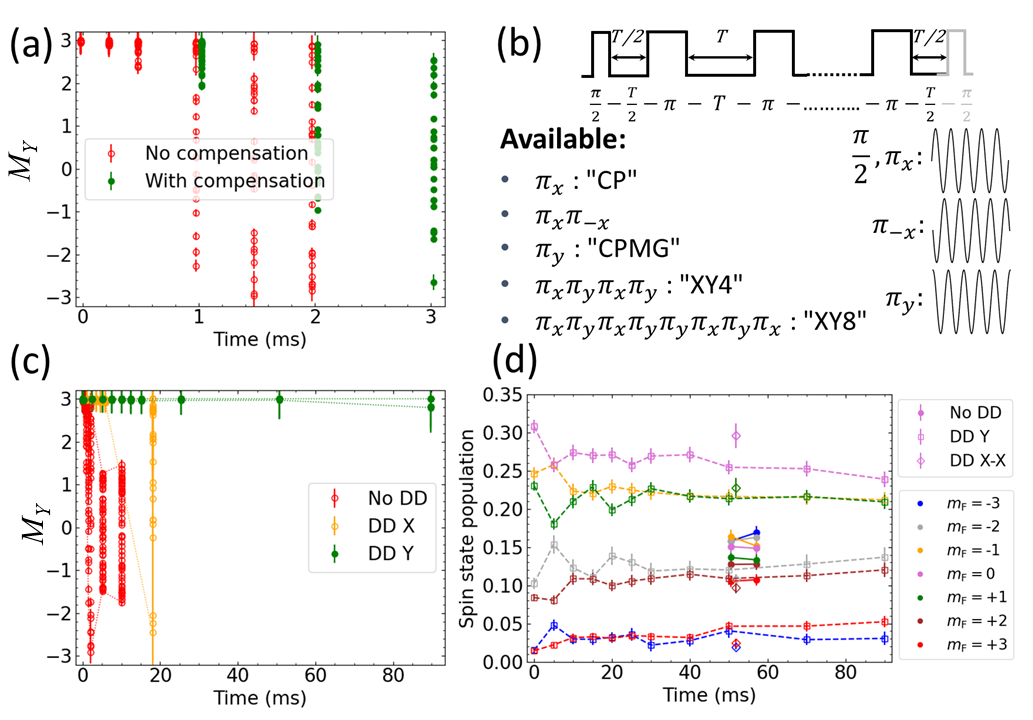}
\caption{\setlength{\baselineskip}{6pt} {\protect\scriptsize (a) Use of active magnetic field compensation and of synchronization with the 50 Hz-main allows to increase by a factor $\simeq 2$ the spin coherence, which is not sufficient to measure magnetization over the full dynamical timescale. (b) Description of the different DD sequences studied. Dynamical Decoupling sequences and their outcome: (c) "X" sequence does not preserve coherence long enough, unlike "Y" sequence; but in (d) it is shown that "Y" (or "X-(-X)") sequence freezes dynamics in deep optical lattices: spin populations do not change after a long time, contrary to the case where no DD is applied (see points at 50 and 55 ms).}}
\label{figSM2}
\end{figure}

\begin{itemize}

  \item  X sequence: pulses are repeatedly applied corresponding to rotation along the X-axis. We find that desynchronization is total after $\simeq 15$ ms (see Fig \ref{figSM2} (c)), which we interpret as sensitivity to non-robustness to RF pulse imperfections.

  \item  Y sequence (also known as CPMG): repeated rotations around the Y axis. Spin dynamics is totally frozen at large lattice depths (see Fig.\ref{figSM2} (d)). As shown below, this arises because an imperfect DD sequence creates a non-negligible mean effective longitudinal magnetic field on the atoms, i.e. parallel to the mean spin direction. This trivially maintains coherence for a very long time (see Fig. \ref{figSM2} (c)).

  \item  X-(-X) sequence:  pulses are repeatedly applied corresponding to rotation along the x axis, with alternating signs. We find that the dynamics is mostly frozen in deep lattices (see Fig. \ref{figSM2} (d)); our interpretation is that pulse imperfections, combined with magnetic inhomogeneities, are responsible for an effective longitudinal RF field (see below).

  \item  XY4 and XY8 sequences: these two sequences that alternate between rotation in the X direction and the Y direction give very favorable results, that are fully compatible with the expectations on both spin dynamics and the phase of the interferogram. Apart from the Y sequence, XY4 and XY8 DD sequences are expected to be best insensitive to pulses imperfections. We did not extensively compare these two sequences; it seems though that an overall phase drift is more systematic and large with an XY4 sequence than with an XY8 one. We therefore concentrated on the so-called XY8 sequence \cite{Ali2013} to take data.

\end{itemize}

\subsection{Perturbative model}

In order to obtain an intuitive understanding for these observations, we derived a perturbative model, where  pulses are separated by free evolution time of duration $dt$. In our framework, pulses are described by a simple rotation of spins, whereas during the free evolution time, the system of two spins ${\vec{ \hat S}}_1$ and ${\vec{\hat S}}_2$ evolves under the effect of the Hamiltonian $\hat{H_{\rm{fe}}}$:

\begin{equation}
    \hat{H}_{\rm{fe}}= \gamma {\vec{\hat S}}_1\cdot \vec{B}_1 + \gamma {\vec{\hat S}}_2\cdot \vec{B}_2 + V_d \left[  \hat S_1^Z \hat S_2^Z - \frac{1}{2} \left( \hat S_1^X \hat S_2^X + \hat S_1^Y \hat S_2^Y \right) \right]
\end{equation}
$H_{\rm{fe}}$ includes interactions between two atoms and interaction with an inhomogeneous magnetic field of values $\vec{B}_1$ and $\vec{B}_2$ at the position of each atom $i$. We will take $\vec{B}_i$ to be along the $Z$ axis (with an amplitude $B^{(i)}_Z$). The free evolution of spin 1 under the effect of $\hat{H}_{\rm{fe}}$ is given by
\begin{equation}
    \frac{d \vec{\hat S}_1}{dt} = \frac{1}{i} \left[ \vec{{\hat S}}_1 , \hat{H}_{\rm{fe}} \right] = -\gamma\vec{\hat S}_1 \times \vec{B}_1  -\gamma\vec{\hat S}_1 \times \vec{B}_{dd}
\end{equation}
where $\vec{B}_{dd} = b_{dd} \left( -\frac{1}{2} \langle\hat S_2^X\rangle , -\frac{1}{2} \langle \hat S_2^Y \rangle , \langle \hat S_2^Z \rangle\right)$.

Starting with spins in state $\vec{S}_{i=(1,2)}(t)$, we allow the evolution due to $H_{fe}$ for a time dt. Then we perform a rotation along a given axis (X or Y). We allow the evolution due to $\hat H_{\rm{fe}}$ for a time dt again, and perform a final rotation along a given axis (X or Y). We define an effective Hamiltonian $H_{\rm{eff}}$ such that the final state of the spin after these 4 steps is $\vec{S}_{(i)}(t+2dt)=\vec{\hat S}_{(i)}(t) + 2 dt/i \left[  \vec{\hat S}_{(i)} , \hat H_{\rm{eff}} \right]$.
We now describe the main simple findings from this approach, keeping only terms that are linear in dt.

\begin{itemize}

\item {\it Perfect $\pi$ rotations along X or Y}-
For $\pi$  rotations around the X axis, which corresponds to $\hat S_X \rightarrow \hat S_X$, $\hat S_Y \rightarrow -\hat  S_Y$, $\hat S_Z \rightarrow - S_Z$, we recover the expected result $\hat H_{\rm{eff}} =  V_d \left[ \hat S_1^Z \hat S_2^Z - \frac{1}{2} \left( \hat S_1^X \hat S_2^X + \hat S_1^Y \hat S_2^Y \right) \right]$. The dynamical decoupling technique removes the sensitivity to magnetic fields, without affecting the dipole-dipole interactions. The same result is obtained for rotations around Y, $\hat S_X \rightarrow - \hat S_X$, $\hat S_Y \rightarrow  \hat S_Y$, $\hat S_Z \rightarrow - \hat S_Z$.

We now consider imperfect rotations.

\item {\it Imperfect rotations along Y}-
We consider the case where imperfect $\pi$  rotations are applied around the Y axis. Each rotation  corresponds to $\hat S_X \rightarrow - \hat S_X - \epsilon \hat S_Z$, $\hat S_Y \rightarrow  \hat S_Y$, $\hat S_Z \rightarrow - \hat S_Z + \epsilon \hat S_X$, we find an effective Hamiltonian:
{\small\begin{equation}
\hat  H_{\rm{eff}} =    V_d \left[ \hat S_1^Z \hat S_2^Z - \frac{1}{2} \left( \hat S_1^X \hat S_2^X + \hat S_1^Y \hat S_2^Y \right) \right] + \gamma \vec{\hat S}_1 \cdot \vec{B}^{(1)}_{\rm{eff}} + \gamma \vec{\hat S}_2 \cdot \vec{B}^{(2)}_{\rm{eff}}
\end{equation}}
with $\vec{B}^{i=1,2}_{\rm{eff}}=  \frac{\epsilon}{ \gamma dt} \vec{Y} $. We therefore obtain that in the case of dynamical decoupling in the Y direction only (Y DD), an inaccuracy in the $\pi$ pulses creates an average mean magnetic field along the magnetization axis. This  finding  explains our experimental results using this DD sequence. Indeed, if this mean magnetic  field, parallel to the atomic spins, defines an energy scale large compared to DDIs, the spins remain in a polarized state (parallel to $\vec{Y}$), and no dynamics takes place.  In practice, we could not reach a sequence where $\vec{B}_{\rm{eff}}$ is kept below the strength of the dipolar interactions, which in turn defines a minimal value for the angular error $\epsilon$: $\epsilon \gg T/ V_d\simeq10^{-5}$ rad, with $T$ the duration of free evolution between two consecutive pulses, see Fig \ref{figSM2}.

\item {\it Imperfect rotations along X}-
We now consider rotations along the X axis; in this case, we alternate between clockwise and anti-clockwise rotations (sequence X-(-)X). This corresponds respectively to ($\hat S_X \rightarrow  \hat S_X $, $\hat S_Y \rightarrow - \hat S_Y - \epsilon \hat S_Z $, $\hat S_Z \rightarrow - \hat S_Z + \epsilon \hat S_Y$) and to ($\hat S_X \rightarrow  \hat S_X $, $\hat S_Y \rightarrow - \hat S_Y + \epsilon \hat S_Z $, $\hat S_Z \rightarrow - \hat S_Z - \epsilon \hat S_Y$). We find an effective Hamiltonian:
{\small\begin{equation}
 \hat{H}_{\rm{eff}} =    V_d \left[ \hat S_1^Z \hat S_2^Z - \frac{1}{2} \left( \hat S_1^X \hat S_2^X + \hat S_1^Y \hat S_2^Y \right) \right] + \gamma \vec{\hat S}_1 \cdot\vec{B}^{(1)}_{\rm{eff}} + \gamma \vec{\hat S}_2\cdot\vec{B}^{(2)}_{\rm{eff}}
\end{equation}}
with $\vec{B}^{(i)}_{\rm{eff}}=  \frac{\epsilon B^{(i)}_z}{2} \vec{Y} $. Therefore the effect of pulse imperfections on such a sequence is to rotate the inhomogeneous magnetic field into the $Y$ direction, $i.e.$ parallel to the spins - which can freeze spin dynamics (as experimentally observed).

\item {\it XY4 sequences}-
We now consider XY4 sequences which alternate imperfect $\pi$ pulses around the Y axis, with an errors $\epsilon$; and imperfect $\pi$ pulses around the X axis, with error $\epsilon '$. This sequence is a more complex sequence, as it is constituted of repetitions of 4 pulses $\pi_X, \pi_Y, \pi_X, \pi_Y$ each separated by a free evolution for a duration dt. We thus define an effective Hamiltonian $\hat{H}_{\rm{eff}}$ such that the final spin state is $\vec{\hat S}_{(i)}(t+4dt)=\vec{S}_{(i)}(t) + 4 dt/i \left[ \vec{\hat S}_{(i)}, \hat H_{\rm{eff}}  \right]$.
We then find that, to first order in $\epsilon$ and $\epsilon'$, the effective Hamiltonian is simply $\hat H_{{\rm{eff}}} =  V_d \left[ \hat S_1^Z \hat S_2^Z - \frac{1}{2} \left( \hat S_1^X \hat S_2^X + \hat S_1^Y \hat S_2^Y \right) \right]$. This indicates that the dynamical decoupling technique is insensitive to pulse errors, and that no transverse mean magnetic field arise, even in the presence of imperfect pulses.

\end{itemize}

\section{IV DD XY 8 performances}

In this section, we analyze the level of performance of the XY8 DD sequence that we have used in this study.

\begin{figure}
\centering
\includegraphics[width= 3.4 in]{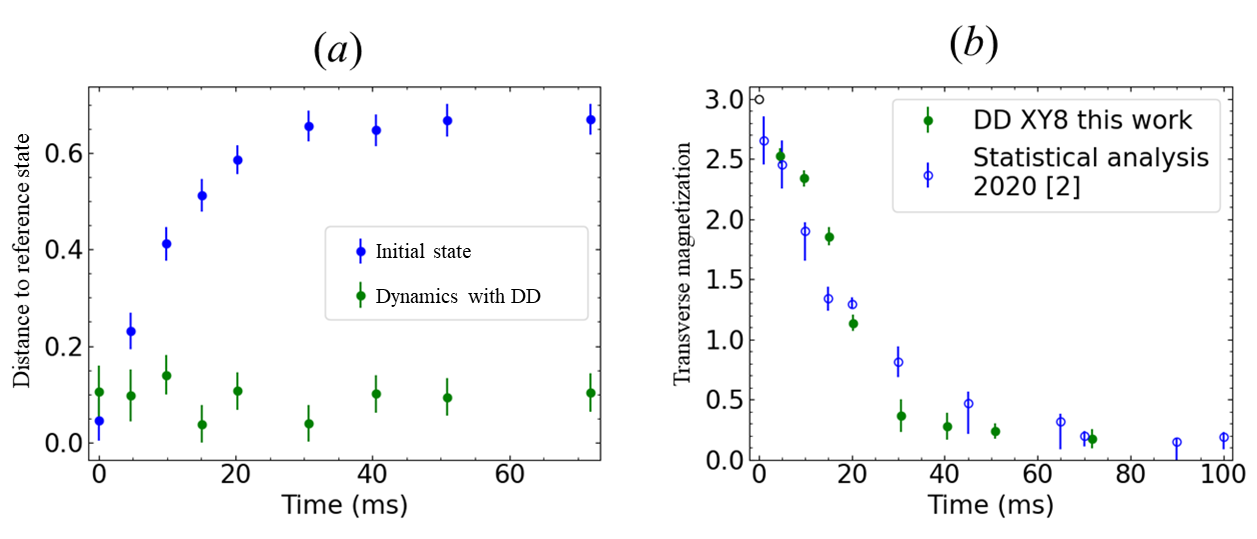}
\caption{\setlength{\baselineskip}{6pt} {\protect\scriptsize Dynamics deep in the Mott insulator regime. a) Comparison of spin population dynamics with or without DD. The distance (see text) between spin populations without applying DD (chosen as the reference) and the spin populations when applying DD remains small (green symbols) compared to the distance with the initial spin state populations (blue). b) Dynamics of transverse magnetization with DD (full green symbols; this work), or with a simple echo and using statistical analysis to infer $M_Y$ (open blue symbols; previous work).}}
\label{figSM3}
\end{figure}

In Figure \ref{figSM3}, we show results to quantify how close is the evolution of the system with DD compared to the one without DD, in large depth lattices. In Fig \ref{figSM3}(a), we quantify the difference between spin populations by evaluating relative distances $\sum_{m_S}\left| P_{m_S,i}-P_{m_S,\rm{ref}}\right| / P_{m_S,\rm{ref}}$, where the reference spin populations $P_{m_S,\rm{ref}}$ correspond to a dynamics with no DD applied. The distance with populations when DD is applied remain small along the whole dynamics, which is compared to the large distance increase with the initial spin state.
In Fig \ref{figSM3}(b), we compare the data we have obtained using DD with previous data (see \cite{Gabardos2020}) for which we were using a simple echo. The results are quite close. The small discrepancies may result from use of deeper lattice in the previous study ($30 E_R$) compared to the actual one ($25 E_R$), which can result in a different value of the tensor light shift that slightly affects the dynamics, see \cite{Lepoutre2019}.

\begin{figure}
\centering
\includegraphics[width= 3.4 in]{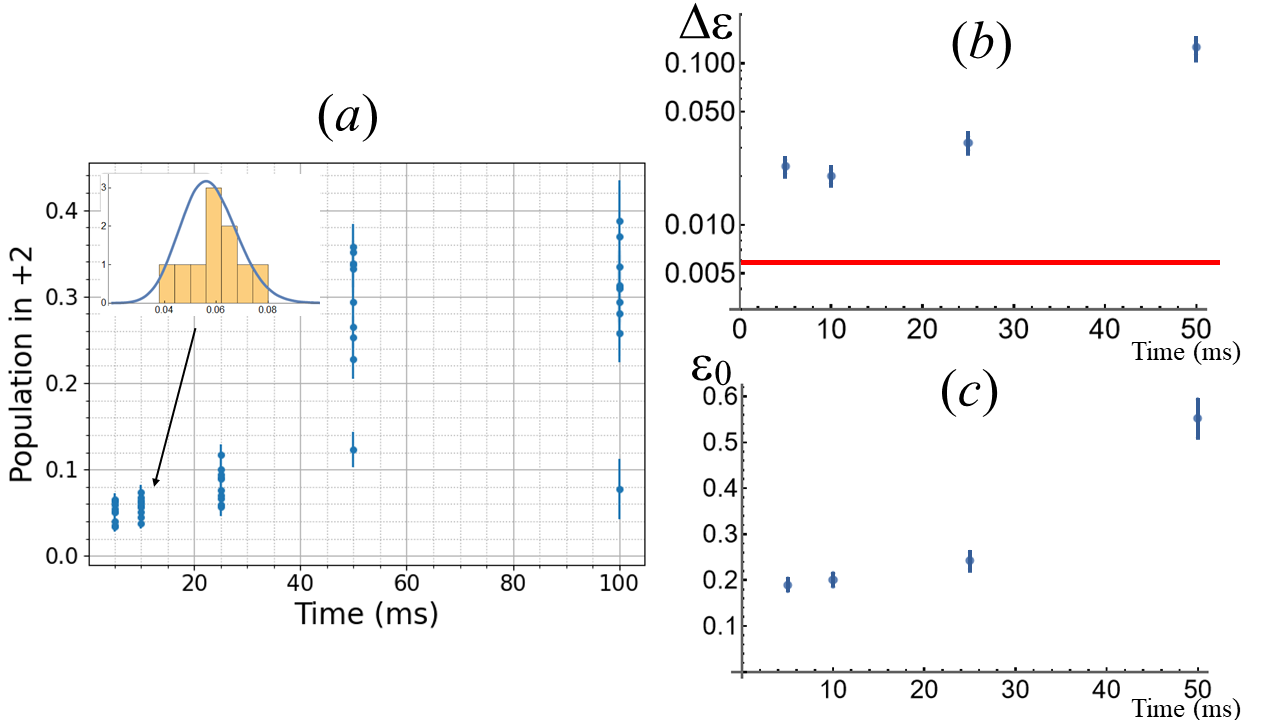}
\caption{\setlength{\baselineskip}{6pt} {\protect\scriptsize Outcome of a Ramsey experiment with XY8 DD as a function of time, in a BEC. (a): Imperfections of DD are characterized by measurement of spin population in $m_s=+2$. Analysis of the distribution of this spin population (see inset) yields the standard deviation of the angle associated to desynchronization $\Delta \epsilon$, which is compared in (b) to the standard quantum noise: (red) full line; and to a spin phase drift, $\epsilon_0$, shown in (c). Angles are in radian. Error bars in (b) results mostly from subsampling, while in (c) they are evaluated from fitting of the spin population distribution.}}
\label{figSM4}
\end{figure}

We show in Fig \ref{figSM4} an analysis of the capability of DD to maintain spin coherence as dynamics proceeds. Data are taken in the BEC, where no spin dynamics is expected (see main text and Section I). Therefore, the outcome of a Ramsey experiment should ideally be a simple rotation by an angle $\pi$ of the initial spin state. As the initial spin state corresponds to a product state with all atoms in $m_S=-3$, the expected final state is a product state with all atoms in $m_S=+3$: an imperfection of the DD sequence translates into non zero spin populations in other spin states $m_S<+3$. For small imperfections, analysis of population in $m_S=+2$ is sufficient to characterize the imperfection, which in our case applies until roughly $t\leq 50$ ms, see Fig.\ref{figSM4} (a).

We find that, for $t\leq 50$ ms, the measured spin population distributions in $m_S=+2$ are compatible with the result that would be obtained after rotations with mean angle $\pi+\epsilon_0$ and standard error $\Delta \epsilon$, see inset of Fig \ref{figSM4} (a). A perfect DD sequence would lead to $\epsilon_0=0$ and $\Delta \epsilon=0$. Obtaining the best values for $\epsilon_0$ and $\Delta \epsilon$ from the distribution of populations in $m_S=+2$ is straightforward when assuming pure rotations of a polarized atomic sample.

Figure \ref{figSM4}(c) shows $\epsilon_0$ as a function of time: there is a growing overall phase induced by DD, which seems to be non linear with time.
 We attribute this non zero values of $\epsilon_0(t)$ to systematic effects during the DD sequence, which might be due to pulse errors; as $t$ increases, the number of pulses grows linearly; the nonlinear growth remains to be explained. We experimentally checked that imparting an extra rotation (with an angle $-\epsilon_0$) at the end of the Ramsey sequence allows for compensation of this mean rotation, i.e. populations in $m_S=+2$ become as small as for short time ($t=5$ ms). Nevertheless, we did not use this compensation when taking data, because the exact angle for optimal compensation drifts with time, which might be due to drift of the Larmor frequency at the kHz level. For data in lattice, this mean rotation has small impact on measurement of magnetization: for $t\leq 50$ ms, it reduces the magnetization by (at most) $\simeq 1-\cos[0.5]=12\%$, while for large $t$ magnetization is close to 0, see Fig \ref{figSM3} (b).

Figure \ref{figSM4}(b) shows $\Delta \epsilon$ as a function of time. $\Delta \epsilon$ corresponds to residual fluctuations associated with desynchronization, because of a non-perfect DD sequence. The optimal value at short time ($0.02$ rad) for $t=10$ ms corresponds to a desynchronization rate which is $\simeq1000$ times smaller than the one obtained without desynchronization ($\simeq 2 \pi$ in 2 ms, see Fig \ref{figSM2} (a)).

The natural figure of merit for the fluctuation $\Delta \epsilon$ is given by a comparison with the standard quantum noise (SQN) value, which corresponds to squeezing-like measurements, as we now explain. Assuming a collective spin aligned along the Y axis, squeezing analysis relies on measurements of the spin components in the orthogonal (X,Z) plane. When ultimately measuring spin populations along Z (which corresponds to the direction of the external magnetic field, along which Stern-Gerlach measurements are made), one needs to impart \emph{before} measurement a rotation (angle $\theta$) along Y in order to access all spin quadratures in the (XZ) plane. Taking into account desynchronization, set by a random rotation with angle $\phi$ in the (XY) plane, the fluctuations of the measured spin component set by both $\theta$ and $\phi$, is given by:

\begin{equation}
\delta \hat S_{Z,\theta}=\cos{\theta} \delta \hat S_Z +\sin{\theta}\left(\delta  \hat S_X  \cos{\phi}+S_0\sin{\phi}+\delta  \hat S_Y \sin{\phi}\right)
\label{fluctuations}
\end{equation}
with $S_0$ the norm of the collective spin ($S_0=N M_Y$, with $N$ the atom number).
In absence of desynchronization ($\phi=0$), varying $\theta$ allows for measurements of fluctuations of all quadratures. In presence of desynchronization, the measurement of $\delta S_X$ is the most disturbed, and for small $\phi$ one gets, provided $\delta  \hat S_Y \ll S_0$:
\begin{equation}
\delta  \hat S_{Z,\pi /2} \simeq  \delta \hat S_X  +S_0\phi
\label{fluctuations2}
\end{equation}
Therefore, $ \delta S_X \gg S_0\phi$ is required. Fluctuations of a coherent spin state are given by $\delta S_{\rm{coh}}=\sqrt{3N/2}$ for spin 3 particles, so that the criteria to reach measurements at the quantum limit for all quadratures is:
\begin{equation}
\Delta \phi<(6N)^{-1/2}=\rm{SQN}
\label{fluctuations3}
\end{equation}
with $\Delta \phi$ the standard deviation of $\phi$; in our experiment, $\Delta \phi$ results from magnetic fluctuations, i.e. it corresponds to the standard deviation $\Delta \epsilon$ that was estimated above. We show in \ref{figSM4}(b) the value of SQN for $N=5\times 10^3$ (which is the atom number after 10 ms, see section I). At small time we reach $\Delta \epsilon\simeq3$ SQN. Therefore we need to reach an extra factor 10 in suppression of desynchronization in order to be able to perform squeezing analysis. Use of larger Rabi frequencies, and of circular RF polarization in order to cancel non-resonant terms will be investigated.

\section{V Stability analysis in the superfluid regime}

Here we outline the derivation of the stability analysis in the superfluid regime that is used in the main paper. We base our interpretation on the hydrodynamical equation (see main text):
\begin{equation}
\frac{d \vec{\mu} }{dt}=-\vec{\mu} \times \left( -\frac{\hbar}{2 m} \left( \vec{a} \vec{\nabla} \right)\cdot\vec{\mu} -\frac{\hbar}{2 m} \nabla^2 \vec{\mu} + \frac{g_S \mu_B}{\hbar} \vec{B_{dd}}(\vec{r}) \right)
\label{hydro}
\end{equation}

We consider small excitations starting from an originally homogeneous spin texture with all spins pointing almost exactly in the $Y$ direction. However, since we are performing a stability analysis, we also allow for a very small but non vanishing component of the spin projection in the $Z$ direction. We thus look for solutions $\vec{\mu} = \left( f, \sqrt{1-f^2-g^2},g \right) $, with $f \ll 1$ and $g \ll 1 $.  For simplicity, we assumed a one-dimensional geometry with $ g_S \mu_B \vec{B}_{dd} (z)= \phi_{dd} (1-z^2/d^2) \left( -\frac{1}{2} \mu_x,-\frac{1}{2} \mu_y, \mu_z \right)$, such that $g_S \mu_B \vec{\mu}. \vec{B}_{dd} (z) $ is the dipolar mean-field energy. For a qualitative estimate of $\phi_{dd}$, we estimate the dipolar mean-field energy using $\phi_{dd}= \frac{n_0}{3 \hbar}  \mu_0  (g_S \mu_B )^2  S$ \cite{ODell2004eho}, and an estimate of the BEC central density in the Thomas-Fermi regime, $n_0\approx 1.2 \times 10^{20}$ m$^{-3}$. The $(1-z^2/d^2)$ behaviour is also expected for a dipolar BEC in this regime. Since our BEC is three-dimensional, given the complicated nature of the dipolar mean-field, and the simplified one-dimensional model that we use, we take $d$ as an adjustable parameter.

We use Eq.\ref{hydro} with these assumptions, and assume a Gaussian ansatz for the total density $\propto \mathrm{exp} (-z^2/\sigma^2)$. This leads to
\begin{eqnarray}
\frac{df}{dt} &=& \frac{\hbar}{m} \left( \frac{z}{\sigma^2} \frac{d}{dz} - \frac{1}{2} \nabla^2 \right) g - \frac{3}{2} g \phi_{dd} (1-z^2/d^2) \nonumber \\
\frac{dg}{dt} &=&  - \frac{\hbar}{m} \left( \frac{z}{\sigma^2} \frac{d}{dz} - \frac{1}{2} \nabla^2 \right) f \nonumber
\end{eqnarray}
We describe the spinor BEC excitations using the basis of  Hermite polynomials $H_n(z)$. Those Hermite polynomials are the stationary solutions of the ferrofluid hydrodynamic
equation and correspond to a locally polarized gas with spatially varying direction of polarization. Assuming that the spinor state initially projects mostly on $H_0$, we restrict the dynamics to the $H_0$ and $H_2$ modes due to dipolar interactions. We thus set $f= \alpha_0(t) H_0(z/\sigma) + \alpha_2(t) H_2(z/\sigma) $ and $g= \beta_0(t) H_0(z/\sigma) + \beta_2(t) H_2(z/\sigma) $ with $ \alpha_2(0)= \beta_2(0)=0$, in order to perform a stability analysis. $\beta_0 = \beta_0(0)$ is assumed to be small but not vanishing. We find:
\begin{eqnarray}
\dot{\alpha}_0 H_0 + \dot{\alpha}_2 H_2 &=&  2 \omega \beta_2 H_2 - \frac{3}{2} \left( \beta_0 H_0 + \beta_2 H_2\right) B_{dd}(z) \nonumber \\
\dot{\beta}_0 H_0 + \dot{\beta}_2 H_2 &=& - 2 \omega \alpha_2 H_2 \nonumber
\end{eqnarray}

We now use the ortho-normalization property of Hermite polynomials $I_{m,n} \equiv \int dz \mathrm{e}^{-z^2/\sigma^2} H_m(z/\sigma) H_n(z/\sigma) \propto \delta (m,n)$ and define  $\Gamma_{m,n} = \int dz \mathrm{e}^{-z^2/\sigma^2} H_m(z/\sigma) H_n(z/\sigma) (1-z^2/d^2)$. We find:
\begin{eqnarray}
\dot{\alpha}_2 &=&  2 \omega \beta_2 - \frac{3}{2} \phi_{dd} \left( -\beta_0 \frac{\sigma^2}{4 d^2} + \beta_2 \left( 1-\frac{5 \sigma^2}{2 d^2} \right) \right) \nonumber \\
\dot{\beta}_2 &=&  - 2 \omega \alpha_2, \nonumber
\end{eqnarray}
From which the analytical solutions are found:
\begin{eqnarray}
\alpha_2(t) &=& - \frac{\beta_0 C_1 \phi_{dd} \mathrm{sinh}\left( \sqrt{2 C_2 \phi_{dd} \omega - 4 \omega^2} t \right)}{\sqrt{2 C_2 \phi_{dd} \omega - 4 \omega^2}} \nonumber \\
\beta_2(t) &=& \frac{\beta_0 C_1 \phi_{dd} \left( -1 + \mathrm{cosh} \left( \sqrt{2 C_2 \phi_{dd} \omega-4 \omega^2} t\right) \right)}{C_2 \phi_{dd} – 2 \omega} \nonumber
\end{eqnarray}
where $C_1 = -3 \sigma^2 /8 d^2 =3 \Gamma_{0,2}/ 2 I_{2,2}$, $C_2 = 3/2 ( 1-5 \sigma^2 /2 d^2)= 3 \Gamma_{2,2}/ 2 I_{2,2}$. $\beta_0 \neq 0$ accounts for the small but non-vanishing projection of the initial spin state in the $z$ direction.
We note that in the real 3D case, the dipolar interaction has a parabolic saddle shape. Therefore, in some direction $d^2>0$ but in the other $d^2 <0$. As a consequence, the system is unstable both when $\phi_{dd}$ is large and positive and when $\phi_{dd}$ is large and negative. This is in good agreement with the prediction based on a Bogoliubov analysis outlined in \cite{Lepoutre2018csm}. The main added value of the present demonstration compared to \cite{Lepoutre2018csm} is that the trap is naturally taken into account, which turns out to be crucial for the stability analysis.

\section{VI Super-exchange interactions in presence of dipolar interactions}

Before we describe the effect of dipolar interactions on super-exchange interactions and spin dynamics, we first estimate how the presence of doublon-hole excitations due to finite tunneling in the Mott insulating regime. The effect of tunneling in the Mott insulating state can be described using perturbation theory.
We write: $\psi = \psi_{Fock} + \frac{\sqrt{2} J}{U } \sum_{<l,m>} \psi_{l,m} - \frac{J^2}{U^2 } \psi_{Fock} + \frac{J^2}{U^2 } \psi_2 $, where $\psi_{Fock}$ is the Fock state with exactly one atom per site $\psi_{l,m}$ is a Fock state with one atom per site everywhere except in site $l$ where there is no atom, and in site $m$ where there is a doublon. The sum $\sum_{<l,m>}$ is taken over nearest neighbours. The state $\psi_2$  is a sum over configurations that either contain  two holes and two doublons, or contain a hole and a doublon that are not nearest neighbors \cite{Gerbier2005ipa}. We do not explicit $\psi_2$ as it has no impact on spin dynamics at second order. We find that the time dependent transverse magnetization follows:
\begin{equation}
    \left< \hat S_Y \right>(t) = 3 N - \frac{81 t^2}{16} \left( \sum_{i \neq k } \alpha_{i,k}^2 - \frac{2 J^2}{U^2} \sum_{<l,m>} \alpha_{l,m}^2 \right)
    \label{eqpert}
\end{equation}

Since the Mott transition qualitatively occurs for $U/zJ \approx 5.8$ \cite{Jaksch1998cba}, with $z$ the number of nearest neighbors,  the corrections of spin dynamics due to tunneling are very small even close to the Mott transition.  This is in contrast to the effect of tunneling on the momentum distribution revealed by the diffraction peaks, which occurs at first order in $J/U$ \cite{Gerbier2005pco}.

We now turn to the description of the effect of dipole-dipole interactions on super-exchange.  To simplify the discussion, we focus on the case of two $S=1$ atoms in a double-well. We first consider the case where a Fock state is perturbed by weak intersite tunneling $J$, in presence of spin-dependent contact interactions. Then, within second-order perturbation theory, the effective Hamiltonian can be written as
\begin{equation}
\hat{H}_{\rm{no DDI}} = -\frac{J^2}{U_2} \ket{S_t=2} \bra{S_t =2} -\frac{J^2}{U_0} \ket{S_t=0} \bra{S_t =0}
\label{superexchange}
\end{equation}
Where $U_{0,2}$ is the onsite contact interactions associated with the molecular potential $S_t={0,2}$, characterized by a scattering length $a_{0,2}$. Using $1 = \sum_{S_t} \ket{S_t} \bra{S_t}$, $(S_1+S_2)^2 = 4 + 2 S_1.S_2= 2 \ket{S_t=1}\bra{S_t=1} + 6 \ket{S_t=2}\bra{S_t=2} $, and $(S_1+S_2)^4 = 16 + 16 S_1.S_2 + 4 (S_1.S_2)^2 = 4 \ket{S_t=1}\bra{S_t=1} + 36 \ket{S_t=2}\bra{S_t=2} $, one can also write:
\begin{equation}
\hat{H}_{\rm{no DDI}} = c_0 S_1\cdot S_2 +c_1 (S_1 \cdot S_2)^2
\end{equation}
where $c_0$ and $c_1$ can be explicitly written in terms of $J,U_0,U_2$, see \cite{Imambekov2003}.
The situation qualitatively changes in presence of DDIs for three reasons. The first is that DDIs act even in the absence of tunneling (as recalled in the main article). We here focus on two additional impact of dipolar interactions onto super-exchange interactions. These arise because the in-situ interactions (i.e. when two atoms share the same site) depend on the total magnetization of the pair of atoms, and because dipolar interactions couple $S_t=2 ,m_t=0$ to $S_t=0, m_t=0$. Therefore Eq.(\ref{superexchange}) needs to be modified, and reads:
{ \begin{eqnarray}
\hat{H}_{\rm{with DDI}} &=& -\sum_{m_t=(-2,-1,1,2)} \frac{J^2}{U_{2,m_t}} \ket{2,m_t} \bra{2,m_t} \\ \nonumber
&-& \frac{J^2 U_{0,0}}{U_{0,0} U_{2,0}-V_{0,2}^2} \ket{S_t=2,0} \bra{S_t =2,0} \\ \nonumber
&-&\frac{J^2 U_{2,0}}{U_{0,0} U_{2,0}-V_{0,2}^2} \ket{S_t=0,0} \bra{S_t =0,0} \\ \nonumber
&-&  \frac{J^2 V_{0,2} }{- U_{0,0} U_{2,0} + V_{0,2}^2}  \left( \ket{0,0} \bra{2,0} + \ket{2,0} \bra{0,0}  \right)
\label{superexchange2}
\end{eqnarray}
where $ U_{S,m}$ is the in situ interaction expectation energy due to both contact interactions and dipolar interactions for a molecular total spin $S$ and total magnetization $m$, $ U_{S,m}  = U_S +  U_d \bra{S,m} \left[ \hat S_1^Z\hat S_2^Z - \frac{1}{2} \left( \hat S_1^X \hat S_2^X + \hat S_1^Y \hat S_2^Y \right) \right] \ket{S,m} $; and $ V_{0,2} =  U_d \bra{0,0} \left[ \hat S_1^Z \hat S_2^Z - \frac{1}{2} \left( \hat S_1^X \hat S_2^X +\hat  S_1^Y \hat S_2^Y \right) \right] \ket{2,0} $.
The total interaction is the sum of the dipolar Hamiltonian (last term of Eq.~1 in the main article) and $\hat{H}_{\rm{with DDI}}$. We point out that while contact interactions do not lead to appreciable interactions in $S_t=1$, in principle dipolar interactions do. However, because of the symmetric nature of the two-body wavefunction for bosonic particles, on-site interaction with $S_t=1$ necessarily imply band excitations, which can safely be ignored as they correspond to excitation energies much larger than the typical onsite interaction energy scales.

One can then resort to second order perturbation theory in order to discuss the impact of super-exchange interactions on the dynamics in the insulating regime, close to the Mott transition. Starting from a state where both atoms have a magnetization $-1$ along the $Y$ direction, we find:
\begin{equation}
\left<S_Y \right> = \left<S_Y \right> (0) -\frac{t^2}{2} \left( \frac{J^4}{8 U_A^2} + \frac{ J^2 V_{dd}}{8 U_B} – 9 \frac{V_{dd}^2}{8}    \right)
\label{seconorder}
\end{equation}
With

\begin{widetext}

\begin{eqnarray}
\frac{1}{U_A^2} &=& -\frac{6 U_d^2 \left(8 U_2^2 \left(2 U_0^2+U_2^2\right)+4 U_d^3 (5
   U_0+U_2)-2 U_d^2 (U_2-4 U_0) (2 U_0+U_2)+8 U_2
   U_d (U_2-U_0) (2 U_0+U_2)+9 U_d^4\right)}{(U_d-2
   U_2)^2 (U_2+U_d)^2 \left(2 U_0 (U_d-U_2)+U_d^2\right)^2} \nonumber \\
\frac{1}{U_B} &=&  \left(\frac{12 (U_0+U_d)}{U_d^2-2 U_0
   (U_2-U_d)}+\frac{6}{U_2+U_d}\right)
\end{eqnarray}

\end{widetext}

In the limit where the dipolar interactions are small compared to contact interactions, we find simplified expressions:
$\frac{1}{U_A^2} =-\frac{3 (2 U_0^2+U_2^2) U_d^2}{U_0^2 U_2^4}$;
$\frac{1}{U_B} = -\frac{6 U_d  (2 U_0+U_2)}{U_0 U_2^2}$

We cannot quantitatively relate  this discussion to the case of $S=3$ atoms that is described in the main part of the paper. However, in order to provide a qualitative comparison between our experimental data and this toy model, we have evaluated $U_2$, $U_2$ and $U_d$ as a function of lattice depth. To account for the physical parameters of chromium atoms, $U_2$ is supposed to be given by the scattering length in the stretched state $S=6$, and $U_0$ by the scattering length in $S=4$.  Furthermore, in order to take into account the full connectivity, $V_{dd}$ needs to be replaced in the second and third term in parenthesis in the right-hand-side of Eq.~\ref{seconorder} by respectively $\bar{V}$ and $V_{\rm{eff}}$; $\bar{V}=\frac{1}{N}\sum_{i>j}V_{i,j}$ and $V_{\rm{eff}}^2=\frac{1}{N}\sum_{i>j}V_{i,j}^2$ are respectively the average and quadratic average of dipolar interactions. Finally, the spin dynamics rate must be multiplied by $3$ to account for the difference between $S=1$ and $S=3$ particles. See Fig.~3 of the main paper  for a comparison between theory and experiment.

\bibliography{biblioDDArXiv}

\end{document}